\documentclass[twocolumn,showpacs,preprintnumbers,amsmath,amssymb]{revtex4}

\usepackage{graphicx}
\usepackage{dcolumn}
\usepackage{bm}

\newcommand{\MeV}{~\mathrm{MeV}}
\newcommand{\GeV}{~\mathrm{GeV}}
\newcommand{\TeV}{~\mathrm{TeV}}
\newcommand{\zp}{Z^{\prime}}
\newcommand{\uonep}{U(1)^\prime}
\newcommand{\mzp}{M_{\tilde{Z}^\prime}}

\newcommand{\gz}{g_{z^\prime}}

\newcommand{\svev}{\langle S \rangle}

\newcommand{\lams}{\Lambda_S}

\begin{document}


\title{$Z^\prime$-mediated Supersymmetry Breaking}

\author{Paul Langacker$^*$, Gil Paz$^*$, Lian-Tao Wang$^\dagger$, Itay
  Yavin$^\dagger$} 
\affiliation{
$^*$ School of Natural Sciences, Institute for Advanced Study,
  Einstein Drive Princeton, NJ 08540 \\ 
$^\dagger$ Physics Department, Princeton University, Princeton NJ 08544 
}%

\date{\today}

\begin{abstract}
We consider a class of models in which supersymmetry breaking is communicated dominantly via a $\uonep$ gauge
interaction, which also helps solve the $\mu$ problem. Such models can emerge naturally  in top-down constructions and are a version of split supersymmetry. The  spectrum contains heavy sfermions, Higgsinos, exotics, and $\zp \sim 10-100\TeV$; light gauginos $\sim 100-1000\GeV$; a light Higgs $\sim 140\GeV$; and a light  singlino. A specific set of $\uonep$ charges and exotics is analyzed, and we present five benchmark models. Implications for the gluino lifetime, cold dark matter, and the gravitino
and neutrino masses are discussed. 
\end{abstract}

\maketitle

\section{\label{sec:intro}Introduction and Motivation}
To a large extent, the mediation mechanism  of supersymmetry (SUSY)
breaking determines the low energy phenomenology.  A well-studied
scenario is gravity mediation \cite{gravmed}.
During the last
couple of decades, in order to satisfy the increasingly stringent
constraints from flavor changing neutral current measurements, many other
mediation mechanisms, such as anomaly
mediation  \cite{anommed},
gauge mediation  \cite{gaugemed},
and gaugino mediation \cite{gauginomed},
 have been proposed 
(for a review, see~\cite{Chung:2003fi}).  In this letter, we present a
alternative mechanism  in which SUSY breaking is  mediated by  
exotic gauge interactions, such as an additional $\uonep$. Concrete
superstring constructions frequently lead to 
additional, non-anomalous, $\uonep$ factors in the low-energy theory
(see, e.g., \cite{strings})
with properties allowing a $\uonep$-mediated SUSY breaking.
Scenarios with an extra $\uonep$ involved in supersymmetry breaking
mediation have been studied in various contexts \cite{u1med}.
Here, we study  a new
scenario where $\zp$-mediation is the dominant source for both scalar and
gaugino masses.

Another ingredient we would like to consider is the
$\mu$-problem of the Minimal Supersymmetric Standard Model (MSSM). One
class of  solutions invokes a spontaneously broken Peccei-Quinn symmetry 
(see, e.g., ~\cite{Accomando:2006ga}). From the
point of view of top-down  constructions it is common that such a
symmetry is promoted to a $U(1)'$ gauge symmetry~\cite{u1mu}.
Identifying this $\uonep$ with
the mediator of SUSY breaking sets $\mu$ (as well as $\mu B$)
to  the scale of the other soft 
SUSY breaking parameters, which are of the right size  whether or not the
electroweak symmetry breaking is finely tuned.

\begin{figure}
 \begin{center}
 \includegraphics[scale=0.8]{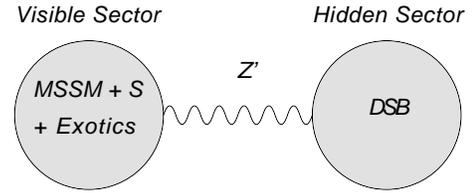}
 \end{center}
 \caption{$Z^\prime$-mediated supersymmetry breaking.}
 \label{fig:two-sectors}
\end{figure}
In the setup we propose, schematically shown in
 Fig.~\ref{fig:two-sectors}, visible and hidden sector fields do not have 
 direct renormalizable coupling with each other. At the same time,
 they are both  charged under $\uonep$.   A supersymmetry breaking
 $\zp$-ino mass term, $\mzp$, is generated due 
to the $\uonep$ coupling to the hidden sector. The observable sector
fields feel the supersymmetry breaking through their couplings to
$\uonep$.
The sfermion masses are of order $m_{\tilde{f}}^2 \sim
\mzp^2/16 \pi^2$. The $SU(3)_C \times SU(2)_L \times U(1)_Y$ gaugino
masses are generated at higher loop order, $M_{1,2,3} \sim  \mzp /(16
\pi^2)^2$, which is 2-3 order of magnitudes lighter than the
sfermions.   LEP direct searches suggest electroweak-ino
masses $> 100$ GeV. We  
therefore expect that  the sfermions are heavy, typically about $100$ TeV. 
In this sense, this scenario can be viewed as a mini-version of
split-supersymmetry~\cite{split}. 
In particular, one fine-tuning is 
needed to maintain a low electroweak scale. This scenario does not have flavor
or CP violation problems due to the decoupling of  the sfermions.
One important difference from split-supersymmetry is the
 $\mu$-parameter, which is set by the scale of $\uonep$ breaking. 

\section{Generic Features of $Z^\prime$-mediated Supersymmetry
  Breaking}
\label{sec:generic}
The visible sector contains an extension of the MSSM. First, we
introduce an extra 
$U(1)^{\prime}$ gauge symmetry. Second, the $\mu$ parameter is promoted
into a dynamical field, $\mu H_u H_d \rightarrow \lambda S H_u
H_d$. $S$ is a Standard Model singlet which is charged under the
$U(1)^{\prime}$. Third, we include exotic
matter multiplets with Yukawa couplings 
to $S$, $\sum_{i \in \{ \mbox{exotics} \}} Y_i {S} X_i X^{c}_i$. They
are included to cancel 
the anomalies associated with the  $U(1)^{\prime}$.
Such exotics and  couplings generically exist in string theory constructions.

\subsection{Features of the Spectrum}

We parameterize the  hidden sector supersymmetry breaking by a
spurion field $X=M + \theta^2 F$. At the scale $\Lambda_S$, supersymmetry
breaking 
is assumed to generate a mass $\mzp \sim
\gz^2(F/M)/16  \pi^2$ for the fermionic component of the
$\tilde{Z}^\prime$ vector superfield.

We assume that all the chiral superfields in the visible sector are
charged under $\uonep$, so all the corresponding scalars receive  soft
mass terms at 1-loop,
\begin{equation} 
\label{eqn:scalarmass}
m^2_{\tilde{f}_i} \sim \frac{\gz^2 Q_{f_i}^2}{16 \pi^2} \mzp^2
\log\left(\frac{\Lambda_S}{\mzp} \right) \sim (100\TeV)^2 
\end{equation}
where $\gz$ is the $\uonep$ gauge coupling and $ Q_{f_i}$ is the $\uonep$
charge of $f_i$, which we take to be of order unity. 

The $SU(3)_C \times SU(2)_L \times U(1)_Y$ gaugino masses
can only be generated at 2-loop level 
since they do not directly couple to the $\uonep$ gaugino,  
\begin{eqnarray}
\label{eqn:gauginomass}
&\quad& \nonumber \\
{M}_a~~ &\sim& \parbox[t]{5cm}{\vspace{-1cm}
  \includegraphics[scale=0.45]{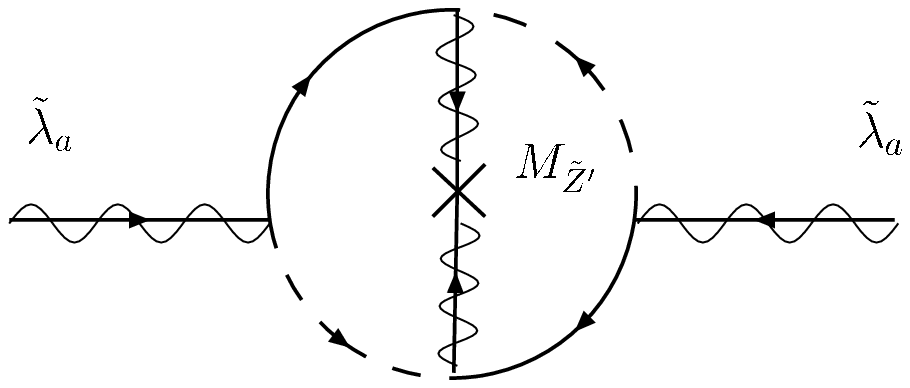} }  \\ \nonumber  
&\sim& ~~\frac{\gz^2 g_a^2}{(16\pi^2)^2} \mzp
\log\left(\frac{\Lambda_S}{\mzp} \right) \sim 10^2-10^3\GeV
\end{eqnarray}
where $g_a$ is the gauge coupling for the gaugino
$\tilde{\lambda}_a$. It is straightforward to verify that this is indeed
the leading $\uonep$ contribution to the  gaugino mass. In particular,
kinetic mixing induced by loops of visible sector fields does not
contribute significantly  due to  chiral symmetries.

The gravitino mass $m_{3/2} \sim F/ M_P$ depends strongly on the scale
of supersymmetry breaking.
Requiring MSSM gaugino masses
$\geq$ 100 GeV and assuming
$\sqrt{F}$, $M$ and $\lams$ to be of the same order of magnitude,
we find  $\sqrt{F} \sim 10^7-10^{11}$ GeV. This is very different
from gauge 
mediated supersymmetry breaking, where the lower 
 scale ($\sim 10 - 1000$ TeV ) typically implies a gravitino
much lighter than the other superpartners. 
Here, the
scale is constrained logarithmically by the
requirement of radiative symmetry breaking. 
Therefore, the gravitino mass is
exponentially sensitive to the choice of model parameters.

We also expect contributions to gaugino masses through gravity
mediation of the order $F/M_P$, which could be of the same order as
Eq.~\ref{eqn:gauginomass}. However, its contribution to scalar masses
$\sim F^2/M_P^2$ is negligible
compared with the
$\zp$-mediation.   Therefore, we expect the hierarchy between scalar
and gaugino masses to be generic.

\subsection{Symmetry breaking and fine-tuning}
\label{sec:symmetry-breaking}

The $\uonep$ gauge symmetry must be broken by the  singlet's VEV $\langle S
\rangle$. We assume this 
is triggered by radiative corrections  to the soft mass $m_S^2$,
especially through Yukawa couplings to exotics. 
Therefore, successful radiative
breaking of $\uonep$ usually requires that those couplings are not small.
$\svev$ is parametrically only an order of magnitude
smaller than $\mzp$. 
It is therefore reasonable to first determine $\svev$ ignoring the
Higgs doublets, and then to consider the Higgs potential for the
doublets regarding 
$\svev$ as  fixed. 

To generate the electroweak scale $\Lambda_{\rm EW}$ we must fine-tune one
linear combination of the two Higgs doublets to be much lighter than
its natural scale. 
The full mass matrix for the two Higgs doublets is,  
\begin{eqnarray}
\label{eqn:higgsMatrix}
\mathcal{M}_H^2 &=& \left( 
\begin{array}{cc}
m_{2}^2 & -
A_H \svev \\ 
\\
- A_H \svev & m_{1}^2 
\end{array}
\right) \nonumber \\
m_{2}^2 &=& m_{H_u}^2 + \gz^2 Q_S Q_2 \svev^2 + \lambda^2 \svev^2 
\nonumber \\
m_{1}^2 &=& m_{H_d}^2 + \gz^2 Q_S Q_1 \svev^2 +
\lambda^2 \svev^2 .
\end{eqnarray}
Generically, one can tune various elements in $\mathcal{M}_H^2$
to obtain one small eigenvalue $\sim \Lambda^2_{\rm
  EW}$. The up-type Higgs mass term can be driven small or  negative
due to the  
large top Yukawa coupling. One typically finds solutions by tuning 
$|m^{2}_{2}| \ll m_1^2 \sim \gz^2 \mzp^2/16 \pi^2$.  The trilinear
term is smaller, $A_H \sim \lambda \gz^2\mzp/16 \pi^2\sim \lambda
\times 10$ TeV, so  integrating out $H_d$
will not shift the smaller eigenvalue significantly. 
$\tan\beta$ is well
approximated by $ \tan\beta = m_{1}^2 / A_H \svev \sim 10-100 $.
There is a single Standard Model-like Higgs scalar, with mass in
the range $140\GeV$.
The remaining Higgs particles are at a scale of order $\sim 100\TeV$. The Higgs
mass is somewhat heavier than the typical prediction of the MSSM, due
to the $\uonep$ $D$ term and the running of the effective quartic
coupling from $\mzp$ down to the electroweak scale. 

It is possible to tune with all the parameters, such as $\gz$ and
$\lambda$, of the same order. In addition, there is an interesting limit
when $\gz \ll \lambda$. Generically, we expect $\svev \sim \mzp/4
\pi$. The singlino mass is $\sim\gz^2 Q_S^2 \svev^2 / \mzp  \sim \gz^2
\mzp /16 \pi^2 \ll \mzp $. Moreover, since $|m_{H_u}^2|
\propto \gz^2 \mzp^2 /16 
\pi^2$,  to fine-tune $m_2^2 \sim 
\Lambda_{\rm EW}^2$ we expect the parameters  to be
chosen so that the singlet's VEV is even smaller $\svev \sim 
(\gz/ \lambda) \mzp/4 \pi$. Therefore, it is  possible to have the
singlino be very light $m_{\tilde{S}} \sim
\left(10^{-3}-10^{-5}\right) \mzp $.  In certain cases, the $\zp$
gauge-boson, $M_{\zp} \sim \gz Q_S \svev$, could even be  light enough to
be produced at the LHC. 

\section{Model Parameters and Low-Energy Spectrum}\label{sec:model}

The free parameters are   $\gz$,
$\lambda$, the exotic Yukawa couplings, the $\uonep$ charges,
$\mzp$, and the supersymmetry
breaking scale $\lams$. 
The charges are chosen to cancel all the anomalies. A minimal choice,
which also leads to a light wino ($M_2 < M_{1,3}$),
involves the introduction of 3 families of colored exotics ($D$) and two
uncolored $SU(2)$-singlet families ($E$). Normalizing the down-type Higgs
charge to unity, 
$Q_1 = 1$, we are left with two independent parameters, which we
choose to be the up-type Higgs and the left-handed quark charges,
$Q_2$ and $Q_Q$ respectively.  
Several additional constraints need to be satisfied by the choices of
charges and other parameters. $\uonep$ has 
to be spontaneously broken by radiative corrections. It must allow
appropriate fine-tuning to break the electroweak symmetry. Moreover,
since $\uonep$ $D$-terms could contribute to scalar masses with either
sign, one must check for the existence of charge or color breaking minima.  
\begin{table}
\begin{tabular}{|c|c|c|c|c|c|}
\hline
& 1 & 2 &  3 & 4  & 5\\
\hline $Q_2$ & $-\frac{1}{4}$& $-\frac{1}{4}$ & $-\frac{1}{4}$  &  $-\frac{1}{2}$&
$-\frac{1}{2}$ \\ 
 $Q_Q$ & $-\frac{1}{3} $& $-\frac{1}{3}$ &  $-\frac{1}{3}$ & $-2$ &  $-2$ \\
$\gz$ & 0.45  & 0.23 & 0.23  & 0.06& 0.04\\ 
$\lambda$ & 0.5 & 0.8 & 0.8 & 0.3 & 0.3\\ 
$Y_D$ & 0.6 &0.7 & 0.8 & 0.4 & 0.6\\ 
$Y_E$ & 0.6& 0.6 & 0.6 & 0.1 & 0.1 \\
\hline
$\svev$ & $2\times 10^5$ & $7 \times 10^4 $ & $6 \times 10^4$ & $2
\times 10^5$ & $ 8\times 10^4$ \\ 
$\tan\beta$ & 20 &29 & 33 &45 & 60\\ 
$M_1$ & 2700 & 735 & 650 & 760 & 270\\ 
$M_2$ & 710 & 195 & 180 & 340 & 123\\ 
$M_3$ & 4300 & 1200 & 1100 &540 & 200\\ 
$m_H$ & 140  & 140  & 140 &140 & 140 \\ 
$m_{\tilde Q_3}$ &$10^5$ & $ 5 \times 10^4 $ & $4 \times
10^4$ & $ 8\times 10^4$&
$4\times 10^4$ \\  
$m_{\tilde L_3}$ &$ 3 \times 10^5$ & $10^5$ &$10^5$ & $2 \times10^4$ &
$10^5$ \\ 
$m_{3/2}$ &  890 & 3600 & 810 & 3 & 0.1\\ 
$m_{\tilde{S}}$ & 4300 & 230 & 160 & 31 & 4\\ 
$m_{Z'}$ & $ 7 \times 10^4$& $1.5 \times 10^4$& $1.3 \times 10^4$ &
5600 & 2100 \\ 
\hline
\end{tabular}
\caption{ \label{benchmark} Model inputs and superpartner spectrum of
  five representative models.
  Masses are in GeV.  We  fix $\mzp = 10^6$ GeV. The masses of the first two generations of
  squarks and sfermions are typically larger than that of the
  third. The input parameters $\lambda$, $\gz$ and $Y_{D,E}$ are defined
  at $\Lambda_S$. The spectra are calculated using 
  exact Renormalization Group Equations (RGE)  (see, e.g.,
  \cite{Martin:1993zk}). There is a 
  theoretical uncertainty due to multiple RGE thresholds. This mainly
  affects $m_H$, leading to a several GeV uncertainty.  The gravitino mass
  is calculated by $m_{3/2} = \Lambda_S^2/M_P$ assuming $\Lambda_S \sim
  \sqrt{F}$. There could be deviations from this relation
  in some SUSY breaking models which could lead to a gravitino mass that is
  different by up to a couple orders of magnitude (typically lower). For
  details, see   \cite{zp-pheno}. }  
\end{table} 

We have found several regions
in the $(Q_Q,Q_2)$ space where a solution satisfies all the
constraints. A detailed scan will be presented in a forthcoming
publication \cite{zp-pheno}. The results exhibit a variety of
patterns for the low energy spectrum. 
In Table~\ref{benchmark}, we display five  representative models.
Different 
ordering of the MSSM gaugino and singlino masses could give rise to very
different phenomenology. The singlino mass typically has more
variation since it is 
determined by fine-tuning. The appearance of a light $\zp$ in the
spectrum, shown in model~5 (with $\sigma \times $BR$(\zp \rightarrow
\ell \bar{\ell})$ $\gtrsim 10$ fb), could result in a spectacular
signal and help untangle the underlying model. This generically
happens in the case where the singlino is very light.

A wino as the lightest supersymmetric particle (LSP) and its nearly
degenerate charged partner (the degeneracy 
is lifted at one-loop by about $160\MeV$ \cite{Pierce:1996zz} and
allows the decay 
$\tilde W^+ \rightarrow \tilde W^0 +\pi^+$, which results in a
4 cm displaced vertex) have been 
studied extensively \cite{winodom}, especially
in connection with anomaly mediated 
models \cite{anommed}. It can annihilate
efficiently into gauge bosons. For pure thermal production
the dark matter density is too low for the several hundred GeV mass range we
have assumed. However, it can be considerably larger for non-standard
cosmological scenarios.

Due to small mixings, at most of the order ${\lambda
  v}/{\mu \tan \beta }$,  the decay chain involving the singlino and wino will
have a long life-time which could result in a displaced vertex.
For example, depending on whether the decay is two or three-body,  the
  life-time for $\tilde{S} \rightarrow h^{(*)} + \tilde{W}$
or $\tilde{W} \rightarrow h^{(*)} + \tilde{S}$ is   
in the range of $10^{-11}-10^{-19}$s.
This could give an interesting signature in case of $M_2 >
M_{\tilde{S}}$, or  $M_{\tilde{S}} > M_2$ if the $\zp$
is light enough and has an appreciable branching ratio for decay into
the singlino.

There is a wide range of
possible gravitino masses, $m_{3/2} \sim 10^{-3} - 10^{4}\GeV$. With
typical assumptions about cosmology, $m_{3/2}$ is strongly
constrained by Big Bang Nucleosynthesis (BBN). If the gravitino is not the LSP,
we typically require either it to be heavy ($>10\TeV$) so it decays
before BBN, or that the reheating temperature is less than about
$10^{5}-10^{7}$ GeV \cite{Kohri:2005wn}. In the case that the  gravitino is the
LSP and the next to lightest supersymmetric particle (NLSP)
 is the wino, we require the gravitino to be
lighter than about $100$ MeV \cite{Feng:2004mt}. It is particularly
problematic when the singlino is the NLSP since its decay to the gravitino is further
suppressed, unless the singlino density is strongly diluted by some
late time entropy generation. We also note that decaying into a light
gravitino, $m_{3/2} \sim $ MeV, is not observable on collider time
scales since the NLSP is neutral.

Since the squarks are heavy the gluino decays off-shell
\cite{split}. Its life-time is very sensitive to $\gz$
and is given by, 
\begin{equation}
\label{eqn:gluino-lifetime}
\tau_{\tilde{g}} = 4\times 10^{-16}\mbox{sec}~ \left( \frac{
  m_{\tilde{Q}}}{10^2\TeV}\right)^4 \left(
\frac{1\TeV}{M_3}\right)^5 ~\propto~ \frac{1}{\gz^6} .
\end{equation}
Even though the life-time is long enough for the gluino to hadronize it is too short to result in a displaced vertex. 

Since the scalars are heavy, one-loop contributions to most flavor observables (such as $b \to s \gamma$ ) are highly
suppressed. There are also two loop contributions to EDM and muon $g-2$. However, those are suppressed as compared with the Split SUSY
scenario \cite{split} since the Higgsinos are heavy and the singlino-wino mixing is small. The exotic matter in this model is
very heavy and does not enter any collider phenomenology. 

\section{Discussion}

In this letter, we discussed the generic feature of supersymmetry
breaking dominantly mediated by an extra $\uonep$. We have used a
$\uonep$ which forbids a $\mu$ 
term. Such a requirement gives additional constraints and predicts
interesting low energy phenomenology, such as the existence of a light
singlino and $\zp$ in various regions of the parameter space. However,
$\zp$-mediation is 
possible in a wider range of $\uonep$ models, such as
$U(1)_{B-L}$. We expect the hierarchy between the soft scalar masses
and the gaugino masses to be generic, although
the detailed pattern of soft terms could be quite
different. Considering $\zp$-mediation in a broader range of models is
certainly worth pursuing. 

The model presented here does not provide a seesaw
mechanism for neutrino mass.  However, in a simple variant the
$\uonep$ symmetry forbids Dirac Yukawa couplings $ Y_\nu  {H}_u {L}
{\nu}^c$ at the renormalizable level, but allows them to be generated
by a higher-dimensional operator \cite{Langacker:1998ut},  
\begin{equation}\label{eqn:nuvariant}
W_\nu=c_\nu \frac{S}{M_P} {H}_u {L} {\nu}^c.
\end{equation}
This naturally yields small Dirac neutrino masses of order $(0.01
c_\nu)$ eV for $\svev = 100$ TeV. 

There are several scenarios for the decays and lifetimes of the heavy
exotic particles \cite{Kang:2007ib} 
and for gauge unification. These depend on the details of the
$U(1)'$ charge assignments, and will be discussed in \cite{zp-pheno}.

\begin{acknowledgments}
We would like to thank Michael Dine, Nathan Seiberg, and Herman
Verlinde for useful discussions. 
The work of L.W. and I.Y. is supported by the National Science
Foundation under Grant No. 0243680 and the Department of Energy under
grant \# DE-FG02-90ER40542. P.L is supported by the Friends of the
IAS and by the NSF grant PHY-0503584. The work of G.P. was supported
in part by the Department of Energy \# DE-FG02-90ER40542 and by the
United States-Israel Bi-national Science Foundation grant \# 2002272.  

\end{acknowledgments}

\end{document}